Super-resolution Radial Fluctuations Enables Polarization-resolved Nonlinear Optical Nanoscopy

MacAulay Harvey[1], Richard Cisek[1], Sarry Al-Turk[2], Harry. E. Ruda[2,3], Laurent Kreplak[4,†], and Danielle Tokarz[1,*]

[1]Department of Chemistry, Saint Mary's University, 923 Robie Street, Halifax, NS, B3H 3C3, Canada

[2]Centre for Advanced Nanotechnology, University of Toronto, 170 College Street, Toronto, Ontario M5S 3E4, Canada

[3]Department of Materials Science and Engineering, University of Toronto, 184 College Street, Toronto, Ontario M5S 3E4, Canada, Department of Electrical and Computer Engineering, University of Toronto, 10 Kings College Road, Toronto, Ontario M5S 3G4, Canada

[4]Department of Physics and Atmospheric Science and School of Biomedical Engineering, Dalhousie University, Halifax, NS, B3H 4J5, Canada

[†]Kreplak@dal.ca

[*]Danielle.Tokarz@SMU.ca

Abstract:

Second harmonic generation microscopy (SHG) is a powerful imaging modality which has found applications in investigating both biological and synthetic nanostructures. Like all optical microscopy techniques, the resolution of SHG is limited to approximately half the wavelength of the excitation light. Because of this several groups have proposed techniques to enable super-resolution SHG imaging. However, these techniques often involve quite complicated optical setups compared to standard SHG microscopes, a major impediment towards more widespread utilization. Here we apply super-resolution radial fluctuations (SRRF), a commonly used technique for super-resolution fluorescence imaging, to enable super-resolution SHG microscopy. By imaging individual nanostructures, we demonstrate that SRRF can provide resolution enhancement of up to 3× compared to a laser scanning SHG microscope, which is comparable to the best resolution enhancement reported in the literature. Additionally, we show that SRRF maintains the polarization dependence of SHG, therefore enabling super-resolution polarization SHG imaging. Finally, we perform SRRF processing on third harmonic generation images to demonstrate the significant potential of SRRF for other super-resolution nonlinear optical microscopy. Importantly, since SRRF can achieve super-resolution purely through image processing, the technique demonstrated here could be used to enhance the resolution of images obtained using a wide variety of nonlinear optical microscopy setups including both laser scanning and widefield configurations.

Introduction:

Second harmonic generation microscopy (SHG) is a powerful nonlinear optical microscopy technique which provides insights into the molecular structure and local organization of samples. This technique has found applications in biological imaging of assemblies of protein nanostructures such as collagen fibrils, myofibrils, and microtubules [1,2], polysaccharide assemblies such as starch particles [3], and in the characterization of synthetic nanostructures such as nanoparticles, nanowires and molecular ferroelectrics [4–6]. Like all optical microscopy techniques, the lateral resolution of SHG is limited by diffraction to around half of the excitation wavelength. This is a major disadvantage for SHG since its main applications are in the characterization of nanostructures or nanostructure assemblies.

Several groups have proposed techniques for super-resolution SHG imaging to overcome the resolution limit. This has involved a variety of different approaches including re-scan microscopy (1.4× resolution enhancement) [7], image scanning microscopy (1.4×) [8,9], utilizing a photonic nanojet for SHG excitation (2.3×) [10], and multifocal structured illumination microscopy (2.3×) [11]. These approaches all require specialized optics for SHG excitation and/or collection beyond what is found in a typical SHG microscope. This poses a major disadvantage as the use of nonstandard microscope setups increase both the financial cost of the microscope, and the technical complexity involved. This presents a high barrier to entry for researchers interested in super-resolution SHG imaging, particularly those interested in applications who may lack the specialized knowledge needed to implement existing techniques.

One technique which has significant potential to overcome these limitations is super-resolution radial fluctuations (SRRF). Unlike the methods described above SRRF achieves super-resolution purely through image processing by assessing the degree of local radial symmetry within each pixel of a diffraction limited image [12]. While SRRF has achieved widespread adoption for super-resolution fluorescence imaging due to its high versatility and low barrier to entry [13] it has not yet been applied for super-resolution SHG imaging.

Here we investigate SRRF as a tool for super-resolution SHG imaging. By imaging individual nanostructures, we show that SRRF is able to achieve a lateral resolution enhancement of ~3.0× compared to a laser scanning nonlinear optical microscope. We additionally show that SRRF maintains the polarization dependence of SHG and can therefore be used to obtain super-resolution polarization-resolved SHG (PSHG) images of collagen fibrils within a tissue sample. Therefore SRRF-PSHG presents exciting new opportunities for obtaining ultrastructural sample characterization, since it provides lateral resolution enhancement exceeding the best previously obtained results but can be easily implemented using any nonlinear optical microscope (including both laser scanning and widefield setups), without any need for complicated modifications to the microscope or to the experimental workflow. Finally, we demonstrate the broad applicability of SRRF for super-resolution nonlinear microscopy by comparing results obtained from SRRF processing of SHG and third harmonic generation (THG) images.

Materials and Methods:

Sample Preparation: Cubic $BaTiO_3$ nanoparticles (100 nm diameter, US Research Nanomaterials Inc.) were deposited between two cover slips (VWR, 16004-348). Collagen fibrils were mechanically extracted from a bovine lateral digital extensor tendon obtained from a local abattoir and deposited on a glass cover slip for imaging as described in [14]. ZnSe nanowires were grown by the vapor phase growth method [15] on a silicon wafer. Once cooled, the wafer is deposited in water, and sonication is applied separating the nanowires from the substrate. Then a drop of a solution containing nanowires was deposited on a cover slip and allowed to dry leaving nanowires on the glass surface, a cover slip is then placed on top. Starch particles were extracted from fresh maize obtained from a local market by crushing with a pestle and mortar in distilled water and centrifuging the resultant suspension at 5000 g for 4 minutes similar to [16]. The purified starch was subsequently diluted, and a drop was deposited on a cover slip in a parafilm spacer, covered with a cover slip, and imaged before the sample dried. ~10 μm thick sections of formalin fixed bovine lateral digital extensor tendon was prepared as described in [17].

Imaging: SHG and THG intensity imaging and polarization-in polarization-out (PIPO) SHG imaging is performed using our homebuilt nonlinear optical microscope and analysis framework previously described [18]. The excitation laser has a wavelength of 1030 nm, pulse duration of ~200 fs, and repetition rate of 5 MHz. In both the SHG intensity and PIPO SHG images of 100 × 100, pixels are obtained with a pixel size of either 85 nm or 190 nm, and a pixel dwell time of 12 μs. The excitation laser power is set such that the maximum intensity per pixel is ~24 counts in each frame. For SHG intensity imaging we utilize either linearly polarized excitation with the angle selected to obtain optimal SHG intensity ($BaTiO_3$, collagen, ZnSe), or circularly polarized excitation (starch). SHG intensity images are saved as stacks of 400-500 individual frames to provide the temporal information needed for SRRF processing.

For PIPO imaging we obtain stacks of 50 frames of SHG intensity (I) at all combinations of 8 polarization angles (θ) in excitation and 8 angles (φ) of a linear polarizing filter in the collection path. At the end of the imaging process a final 50 frames are obtained at the initial values of θ and φ to verify that there is no movement or damage of the sample during imaging, and to provide a resolution reference for SRRF analysis (see below). Both the raw and SRRF processed polarization data is then fit to the following equation using a custom MATLAB (The MathWorks, Inc) program to obtain ρ, a commonly reported structural parameter for PSHG microscopy.

$$I = A|\rho \cos^2(\theta - \delta)\cos(\phi - \delta) + \sin^2(\theta - \delta)\cos(\phi - \delta) + \sin 2(\theta - \delta)\sin(\phi - \delta)|^2 + F \quad (1)$$

Here δ is the angle of the structure under investigation within the image plane with respect to the arbitrarily defined polarization of the excitation laser at θ = 0, A is the constant of proportionality between the square of the SHG field and the number of counts measured on the detector, and F is the noise. This equation is valid for cylindrically symmetric assemblies of uniaxial SHG emitters (e.g. the peptide bonds in a collagen triple helix [19]). A more general symmetry (e.g. trigonal [20]) may yield more details about higher order organization within a sample, however, for the highly aligned collagen in tendons, which will be imaged here, equation (1) should still be applicable.

SRRF Processing: SRRF and Fourier ring correlation (FRC) [21] analysis is performed using code which is publicly available as part of the NanoPyx package [22]. A full description of SRRF, and evaluation of its performance for various fluorescence modalities can be found in [12]. Briefly, SRRF assumes that the image is formed from point sources convolved with a radial symmetric point spread function, and therefore the degree of local radial symmetry (radiality) can be used to enable localization beyond the diffraction limit. To obtain super-resolution each pixel of the image is first split into a user defined number of sub pixels (the magnification parameter), the degree of radial symmetry over a user defined ring radius is then determined at each pixel of the image. Since noise may also lead to large radiality values SRRF is usually performed on several consecutive frames of intensity data, and analysis of the radiality over time is performed to reject features which are uncorrelated in time. Here we use the pairwise product mean of the radiality for SHG intensity images, and the radiality average for PIPO data, as this was found to preserve the polarization dependence of SHG better than other procedures, particularly at polarization states with low SHG intensity. In general, a higher magnification and smaller ring radius will lead to higher resolution, however care must be taken to obtain optimal resolution without inducing artifacts.

Here the optimal magnification and ring radius is determined by performing SRRF analysis separately on the first and second half of the intensity stack for intensity data, or the first and last PIPO image stacks containing 50 frames each (which are obtained at the same polarization conditions). This results in two presumably identical super-resolved images. A FRC is then performed to obtain the correlation between the two images as a function of spatial frequency. The highest frequency at which the correlation is greater than 1/7 is reported as the resolution, which yields results similar to the FWHM of the point spread function [21]. The magnification and ring radius is selected to optimize the resolution obtained using an FRC, and SRRF analysis of the entire intensity stack, or all PIPO polarization states is then performed using those input parameters.

Results and Discussion:

To begin with we attempt to gain a benchmark for the resolution enhancement provided by SRRF for SHG intensity images by imaging individual nanostructures including $BaTiO_3$ nanoparticles (diameter <100 nm), individual collagen fibrils (diameter 30 – 250 nm for extensor fibrils [23]), and ZnSe nanowires (diameter <200 nm [15]). Representative raw SHG intensity images, and SRRF processed images are shown in Fig. 1. In each case the SRRF image shows significant resolution enhancement compared to the raw SHG intensity images. The SRRF parameters used, and the full width half maximum (FWHM) resolution obtained by fitting a Gaussian of a profile of intensity across each structure are summarized in Table. 1. For these structures we obtain an average resolution enhancement of 2.5×, similar to the largest reported resolution enhancement for super-resolution SHG microscopes based on multifocal structured illumination, or using a photonic nanojet for SHG excitation [10,11]. However, in this case SHG imaging is performed using a standard laser scanning nonlinear optical microscope and resolution enhancement is provided purely through image processing.

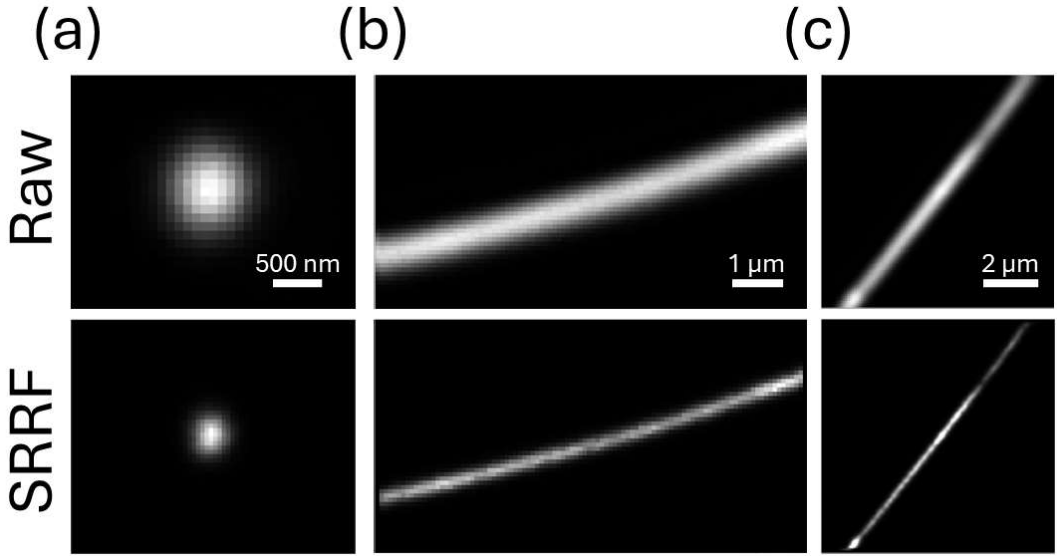

Figure. 1: Demonstration of SRRF resolution enhancement on individual nanostructures. Raw SHG intensity images obtained by summing 400 frames of intensity data, and SRRF processed images of a $BaTiO_3$ nanoparticle (a), a collagen fibril (b), and a ZnSe nanowire (c).

Table. 1: Summary of SHG SRRF analysis of various nanostructures.

| Sample | Ring Radius | Magnification | Raw FWHM | SRRF FWHM | Resolution Enhancement |
|---|---|---|---|---|---|
| $BaTiO_3$ | 2.0 pixels | 2 | 633 nm | 266 nm | 2.4× |
| Collagen | 1.9 pixels | 1 | 548 nm | 255 nm | 2.2× |
| ZnSe | 1.5 pixels | 2 | 525 nm | 173 nm | 3.0× |

To further demonstrate the applicability of SRRF to SHG imaging we perform SRRF analysis of SHG intensity images of larger structures. Within a cluster of multiple $BaTiO_3$ nanoparticles SRRF analysis can clearly resolve individual particles (Fig. 2a). In starch particles SRRF analysis reveals concentric rings of alternating high/low SHG intensity (Fig. 2b). The concentric rings observed are likely starch growth rings composed of alternating semi-crystalline and amorphous shells [24] and are typically not resolved by SHG. Radial channels may also be apparent [25] but are harder to distinguish (Fig. 2b, arrow).

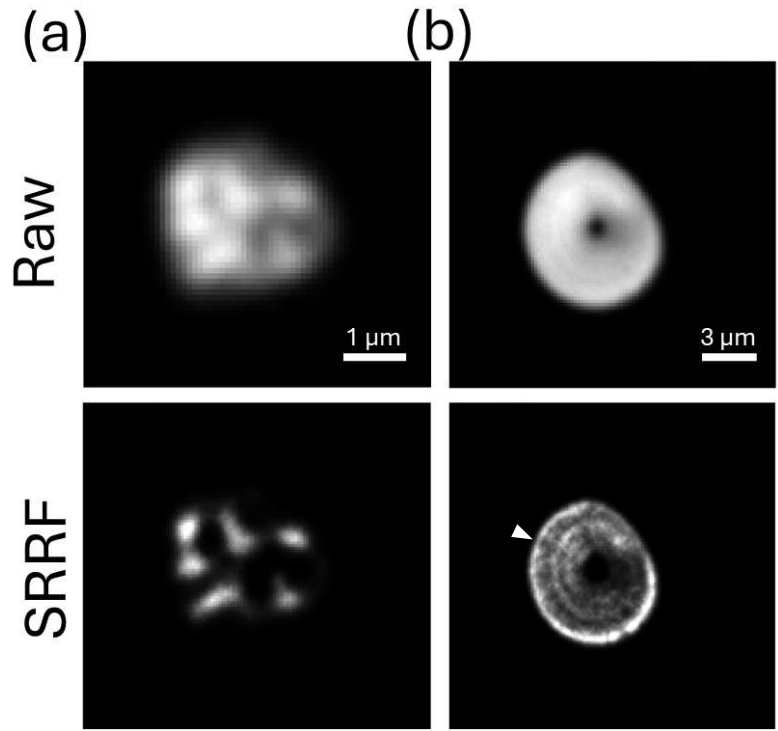

Figure. 2: Demonstration of SRRF resolution enhancement of SHG microscopy data on a cluster of BaTiO$_3$ nanoparticles (a) and a corn starch particle (b). SRRF parameters used are ring radius of 1.7 pixels (a), and 1.5 pixels (b), a magnification of 2 is used in both cases. The resolution of the SRRF processed image obtained from a FRC is 194 nm (a) and 357 nm (b). The white arrow points to a radial channel within the corn starch particle that becomes distinguishable after SRRF analysis.

To demonstrate that SRRF analysis maintains the polarization dependence of SHG we perform PIPO imaging of collagen fibrils within a thin section of bovine tendon. We then perform SRRF processing of the 50 frames of intensity data obtained at each polarization condition using a ring radius of 3 pixels and a magnification of 1. Figure 3a shows the polarization dependence of SHG intensity for a typical pixel from the same location in both the raw and SRRF processed image. These data both fit equation (1) with similar ρ values and high $R^2$ (Fig. 3b), thus demonstrating that SRRF maintains the polarization dependence of SHG. The SRRF intensity image as shown in Fig.4a has a resolution of 316 nm (determined using FRC), this greatly improves the clarity of the SHG intensity image with individual fibrils being very clearly distinguishable. Similarly for the images of the ρ parameter in Fig 4b we see nearly all pixel in the image fit for the raw data, while the SRRF data fits only for pixels located along the individual fibrils. The mean values of ρ are similar between the raw, 1.56 and the SRRF image 1.58, and these values are within the expected range for collagen fibrils [14]. Notably several fibrils in the SRRF images shown in Fig. 4b display a gradient in ρ values perpendicular to the long axis of the fibril. This effect has been previously observed in isolated individual collagen fibrils and occurs because of a phase difference between chiral and achiral contributions to SHG signal from collagen [14] induced at the edge of a high numerical aperture beam focus. The presence of this effect on collagen fibrils within a tissue section is strong evidence that the SRRF algorithm can isolate SHG intensity contributions from individual fibrils. The ability of SRRF enhanced PSHG to visualize individual fibrils within a crowded tissue section will greatly improve the sensitivity of PSHG as a diagnostic tool for a broad range of human pathologies [2,26].

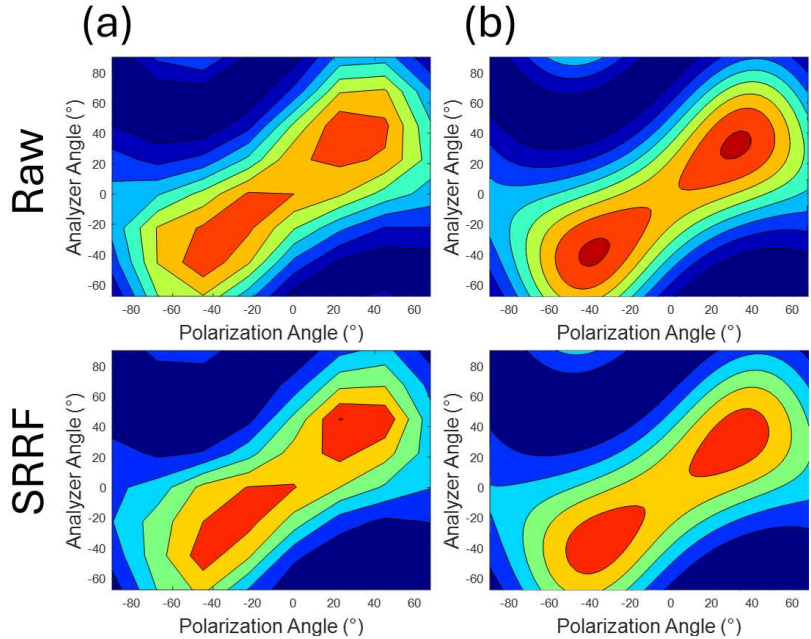

Figure 3: Demonstration that SRRF maintains the polarization dependence of SHG. Contour plots showing experimentally measured polarization dependence of SHG intensity as a function of polarization angle ($\theta$) and analyzer angle ($\phi$) for the same pixel in the raw and SRRF processed PIPO data (a) (color indicates low intensity with blue, high intensity with red). Fits of the data in (a) to equation (1) results in $\rho$ values of 1.49 ($R^2 = 0.94$) for the raw data and 1.54 ($R^2 = 0.93$) for the SRRF processed data (b).

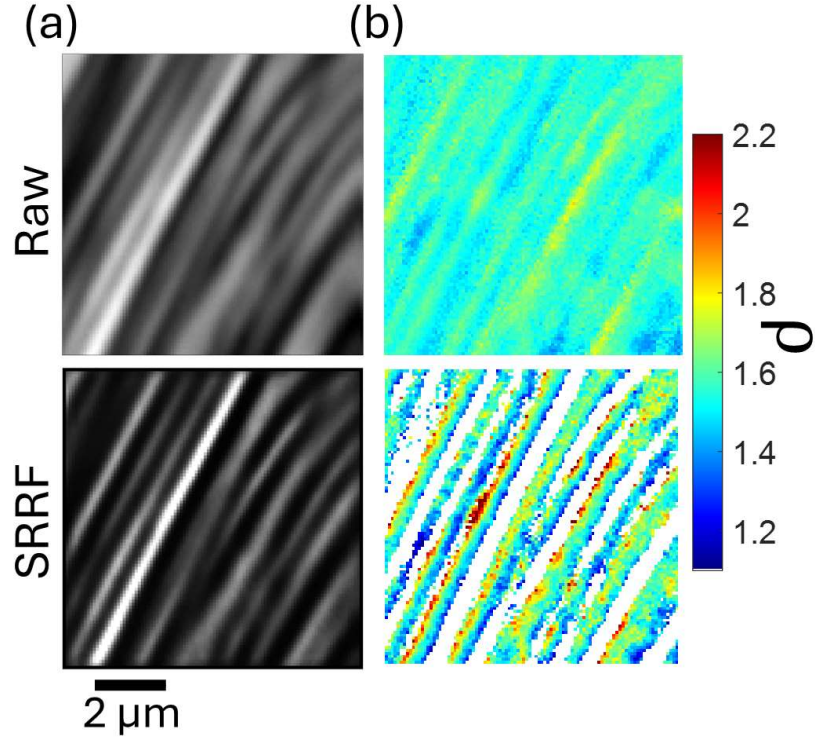

Figure 4: Demonstration of super-resolution PSHG imaging using SRRF. Raw and SRRF processed SHG intensity images obtained by summing SHG intensity from all 64 polarization states used for PIPO (a), fitted ρ values for the raw and SRRF data (b). The FRC resolution of the SRRF image is 316 nm.

In principle, the application of SRRF for super-resolution nonlinear optical imaging as we have demonstrated here for SHG microscopy could also be applied to other nonlinear optical modalities. To show this we perform SRRF analysis of SHG intensity (Fig. 5a) and THG intensity (Fig. 5b) images of a ZnSe nanowire. By fitting a Gaussian function to the profile of intensity across each image (Fig. 5c) we obtain a FWHM resolution of 526 nm for the raw SHG image, 435 nm for the raw THG image, 188 nm for the SHG SRRF image, and 164 nm for the THG SRRF image. Note that the raw FWHM resolutions differ by a factor of $\sim\sqrt{1.5}$ as expected [27]. These results demonstrate that SRRF is able to provide similar resolution enhancement for both SHG and THG images. The THG resolution enhancement obtained here is similar to that recently obtained using THG deactivation microscopy [28]. The application of SRRF to other nonlinear optical modalities such as coherent anti Stokes Raman scattering, or stimulated Raman scattering microscopy, is therefore an exciting area for potential future work.

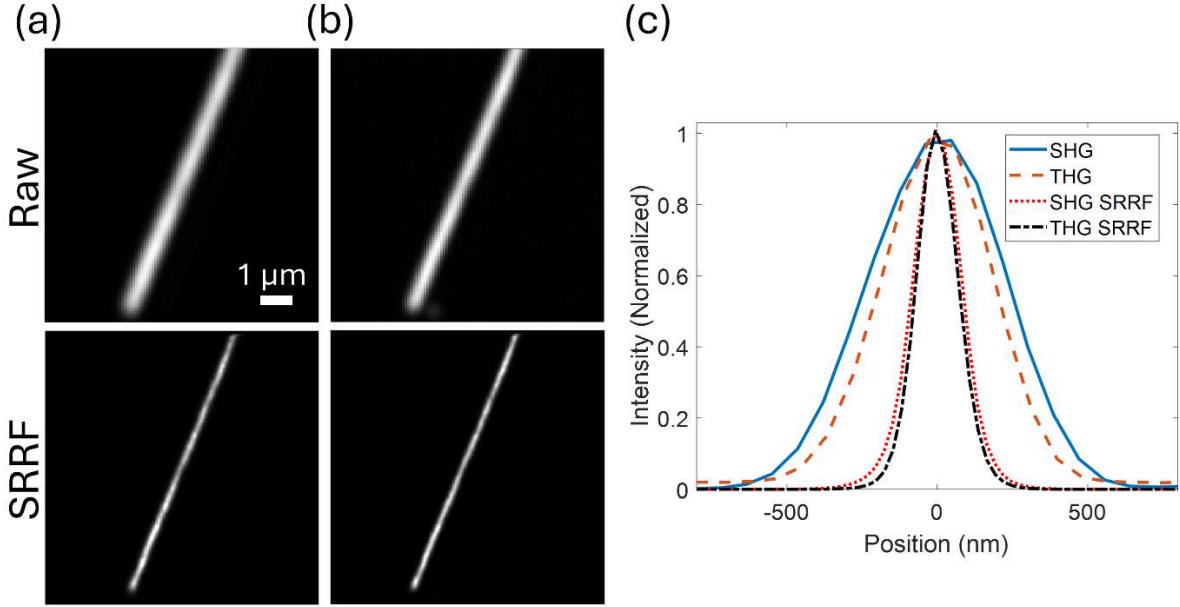

Figure. 5: Application of SRRF for super-resolution THG imaging. Raw and SRRF processed SHG (a) and THG (b) intensity images of a ZnSe nanowire, and plots showing the intensity profile across each image (c). In each case SRRF analysis was performed using a ring radius of 1.6 pixels and a magnification of 3.

The only major disadvantage of SRRF for super-resolution nonlinear optical imaging is that SRRF does not enhance axial resolution, while other techniques have shown axial resolution enhancement of up to 1.5× [11]. However, other radial fluctuation based super-resolution techniques (e.g. eSRRF [29]) can enhance the axial resolution of volumetric data. Therefore, the use of multiphoton techniques such as SHG, THG, or multiphoton fluorescence [30], which already have high axial resolution compared to single photon imaging techniques, for super-resolution volumetric imaging presents an exciting opportunity for development.

Conclusions:

We have investigated the application SRRF image processing for super-resolution PSHG imaging. Through imaging individual nanostructures, we find that SRRF can provide resolution enhancement exceeding the best previously reported super-resolution SHG microscopes. However, since SRRF achieves super-resolution purely through image processing, this technique requires no complicated modifications to the microscope and could in principle be used for super-resolution imaging with almost any nonlinear optical microscope. We additionally demonstrate that the polarization dependence of SHG is maintained by SRRF processing, thereby allowing super-resolution PSHG imaging. Finally, we demonstrate super-resolution THG microscopy using SRRF. These results demonstrate that SRRF used in conjunction with nonlinear microscopy could be a powerful new tool for structural characterization of individual nanostructures and assemblies of nanostructures.

**Funding:** This work was financially supported by Natural Sciences and Engineering Research Council of Canada (NSERC) Discovery Grant funding (RGPIN-2024-04192 for LK and RGPIN-2018-05444 for DT), Canada Foundation for Innovation (John R. Evans Leaders Fund #37749), Research Nova Scotia (1868), Canada's Research Support Fund, and Saint Mary's University. MH is supported by a Nova Scotia Graduate Scholarship and a Scotia Scholars award (Doctoral),

**Disclosures:** The authors declare no conflicts of interest.

**Acknowledgements:** The authors thank Dr. Sam Veres (Saint Mary's University) for his assistance in preparing the tendon sections used here.

**Data Availability:** The data underlying the results presented in this paper are not publicly available at this time but may be obtained from the authors upon reasonable request.